\documentclass[conference]{IEEEtran} 
\IEEEoverridecommandlockouts
\usepackage{cite}
\usepackage{amsmath,amssymb,amsfonts}

\usepackage{algorithmic}
\usepackage{graphicx}
\usepackage{textcomp}
\usepackage{xcolor}
\usepackage{url}
\usepackage{amsmath}
\usepackage{amsfonts} 
\usepackage{amsthm}
\usepackage{clrscode}   

\newtheorem{CodeListing}{Code Listing}

\def\BibTeX{{\rm B\kern-.05em{\sc i\kern-.025em b}\kern-.08em
    T\kern-.1667em\lower.7ex\hbox{E}\kern-.125emX}} 
\begin{document}

\title{Improving the Graph Challenge Reference Implementation
\thanks{Research was sponsored by the Department of the Air Force Artificial 
Intelligence Accelerator and was accomplished under Cooperative Agreement 
Number FA8750-19-2-1000. The views and conclusions contained in this document 
are those of the authors and should not be interpreted as representing the 
official policies, either expressed or implied, of the Department of the Air 
Force or the U.S. Government. The U.S. Government is authorized to reproduce 
and distribute reprints for Government purposes notwithstanding any copyright 
notation herein.}
}

\author{\IEEEauthorblockN{
Inna Voloshchuk, Hayden Jananthan, Chansup Byun, Jeremy Kepner
\\
\IEEEauthorblockA{
MIT
}}}

\IEEEoverridecommandlockouts
\IEEEpubid{\makebox[\columnwidth]{979-8-3315-5937-3/25/\$31.00 ©2025 IEEE \hfill} \hspace{\columnsep}\makebox[\columnwidth]{ }}

\maketitle
\IEEEpubidadjcol

\begin{abstract}

The MIT/IEEE/Amazon Graph Challenge provides a venue for individuals and teams to showcase new innovations in large-scale graph and sparse data analysis. The Anonymized Network Sensing Graph Challenge processes over 100 billion network packets to construct privacy-preserving traffic matrices, with a GraphBLAS reference implementation demonstrating how hypersparse matrices can be applied to this problem. 
This work presents a refactoring and benchmarking of a section of the reference code to improve clarity, adaptability, and performance. The original Python implementation spanning approximately 1000 lines across 3 files has been streamlined to 325 lines across two focused modules, achieving a 67\% reduction in code size while maintaining full functionality.  Using pMatlab and pPython distributed array programming libraries, the addition of parallel maps allowed for parallel benchmarking of the data. Scalable performance is demonstrated for large-scale summation and analysis of traffic matrices. The resulting implementation increases the potential impact of the Graph Challenge by providing a clear and efficient foundation for participants. 
\end{abstract}


\section{Introduction}

The MIT/IEEE/Amazon Graph Challenge is a venue for advancing the state-of-the-art in graph and sparse data analysis.  The Graph Challenge supports novel approaches driven by the community in software, hardware, algorithms and systems to process large graphs and datasets that reflect real-world complexity and scale \cite{jananthan2024anonymized}. 

\begin{figure}
\center{\includegraphics[width=1.0\columnwidth]{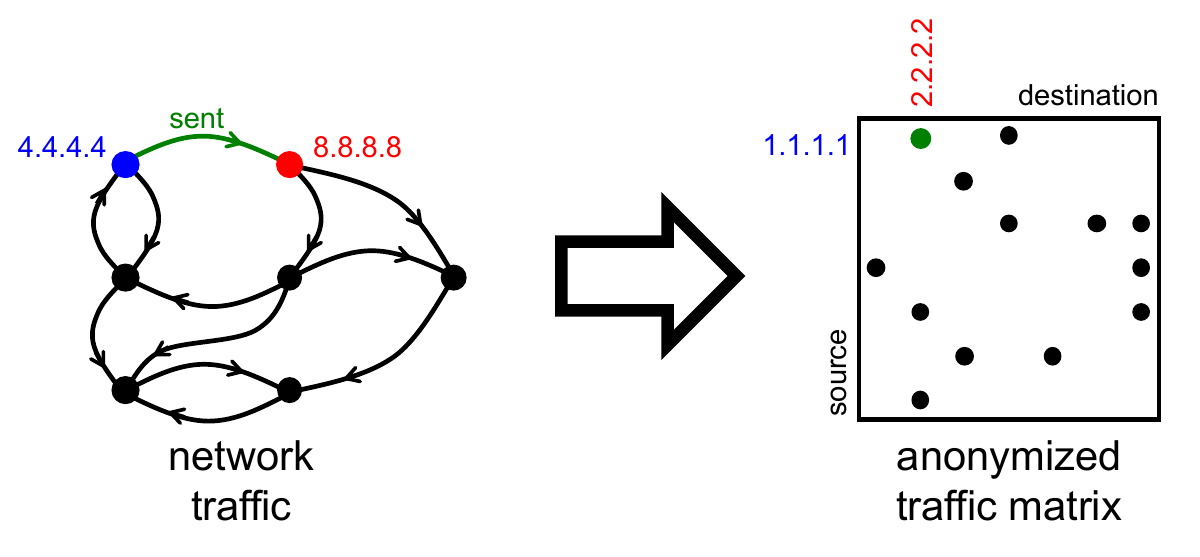}}
      	\caption{{\bf Network Graph to Anonymized Traffic Matrix.} Each network packet contains a source and destination IP address. These addresses can be aggregated into a matrix representation, where rows correspond to sources and columns correspond to destinations. The Anonymized Network Traffic Graph Challenge reference implementation that was modified focused on reading and analyzing traffic matrices made from network traffic. Graphic from \cite{jananthan2024anonymized}.}
      	\label{fig:NetworkToMatrix}
\end{figure}

The Anonymized Network Sensing Challenge is a challenge released within this broader initiative, with a focus on privacy-conscious analysis of network traffic data.  The goal of the challenge is to improve the construction and analysis of anonymized traffic matrices.  By constructing anonymized traffic matrices from streamed network packets, researchers can extract network properties and statistics while preserving privacy \cite{jananthan2024anonymized}.  The traffic matrix representation, described in Figure~\ref{fig:NetworkToMatrix}, enables efficient computation of network properties using linear algebra. 

Network traffic analysis has become increasingly important for cybersecurity and anomaly detection as systems continue to grow in complexity.  The challenge uses more than 100 billion network packets collected by an Internet Telescope operated by the Center for Applied Internet Data Analysis (CAIDA) \cite{CAIDA2022}.  To handle the scale of the data, the provided reference implementation utilizes GraphBLAS hypersparse matrices, which allow no down-sampling, high performance, and open standards \cite{kepner16mathematical, buluc17design,yang2018implementing, kepner2018mathematics, davis2019algorithm, mattson2019lagraph, cailliau2019redisgraph, davis2019write, aznaveh2020parallel, brock2021introduction, pelletier2021graphblas, jones2022graphblas, trigg2022hypersparse, davis2023algorithm}.  This initial reference implementation, while functionally complete, presents refactoring opportunities to increase clarity, conciseness, and performance. 

This work presents a new streamlined reference implementation that reduces code size by 67\% while maintaining functionality. Furthermore, both the new Python implementation and existing Matlab/Octave implementation lend themselves to distributed arrays parallel programming via  pPython \cite{byun2023ppython} and pMatlab \cite{kepner2009parallel}. Benchmarking was completed on all the data, testing different number of threads per process, as well as multi-node scaling up to 32 nodes.

The remainder of this paper is organized as follows. A background is first provided on the Graph Challenge and the specific requirements of the Anonymized Network Sensing Challenge. Then, more detail is given on the specific choices made and changes to the reference implementation. Next, distributed arrays are described, as well as how they were applied to this problem. Finally, benchmarking results are shown, as well as their implications.

\section{Anonymized Network Sensing Graph Challenge}

Network traffic can be represented as a matrix in which each row is a source and each column is a destination (Figure~\ref{fig:NetworkToMatrix}). The data stream of packets is partitioned into time windows, with each time window containing $2^{30}$ packets.  Matrix source and destinations come directly from packet headers, and are able to be anonymized without losing the essence of the network because permutations in rows or columns does not effect the network properties that can be extracted using matrix operations. More information on traffic matrix construction can be found in the Anonymized Network Sensing Graph Challenge Paper\cite{jananthan2024anonymized}. 
 
The Anonymized Network Sensing Graph Challenge consists of six steps:
\begin{enumerate}
    \item Reading packet capture (PCAP) files. 
    \item Extracting source and destination internet protocol (IP) addresses. 
    \item Anonymizing the addresses. 
    \item Constructing traffic matrices. 
    \item Saving matrices into files. 
    \item Reading, summing, and analyzing the stored files.
\end{enumerate}

\begin{figure}
\noindent \rule{\columnwidth}{0.5pt}

{\small\sf
\noindent ReadSumAnalyzeMatrices(
  
~  Np, \# packets in file ($2^{30}$)

~  Nv, \# packets per matrix ($2^{17}$)
  
~  NmatPerFile, \# matrices per output file ($2^{6}$)

\noindent  );
  
  ${\bf A}_t$(:,:) = 0;

  {\bf for} i = 0 {\bf to} (Np/(NmatPerFile*Nv))-1
    
     ~ ${\bf A}$ = readMatrices(i);
    
     ~ {\bf for} j = 0 {\bf to} NmatPerFile-1

     ~  ~ ${\bf A}_t$ += ${\bf A}$[j];
     
     ~ {\bf end}
 
  {\bf end}

  \# perform analysis on ${\bf A}_t$ 

\noindent {\bf end}}

\noindent \rule{\columnwidth}{0.5pt}

      	\caption{{\bf Anonymized Network Sensing Graph Challenge Read, Sum, Analyze Pseudocode}.  Traffic matrix ${\bf A}$[j] is previously created using anonymized source and destination IP addresses. Traffic matrices in groups of $2^6$ are then saved as individual files within a .tar archive.  There are $2^7$ .tar archives saved for each file.  These .tar files are read, and all the traffic matrices are summed into one traffic matrix ${\bf A}_t$.  The analysis in Table 1 found in the Anonymized Network Sensing Graph Challenge Paper \cite{jananthan2024anonymized} is then performed on ${\bf A}_t$ and reported.}
      	\label{fig:AnonNetSenseCode}
\end{figure}

This work focuses on step 6), where $2^{13}$ traffic matrices, representing $2^{30}$ packets, are aggregated into a single matrix ${\bf A}_t$ for analysis as described in Table 1 found in the Anonymized Network Sensing Graph Challenge Paper \cite{jananthan2024anonymized}. Figure~\ref{fig:AnonNetSenseCode} illustrates the reference pseudocode for reading, summing, and analyzing anonymized matrices.

Subrange analysis can also be performed on traffic matrices by using a diagonal matrix as a mask to select a subset of source and destination addresses. Traffic matrices that are created by applying these masks are saved as separate tar archives by the reference implementations for later use.

\section{Implementation Improvements} 

The original Python reference implementation spanned approximately 1000 lines of code, distributed across three files, incorporating features such as hierarchical updates, extended statistical options, and object-oriented abstractions through a {\tt TrafficMatrix} class. While these features provided flexibility for more use cases, they introduced complexity that obstructed the primary goal of providing a clear, accessible reference implementation. 

\begin{figure}
    \centering
    \includegraphics[width=1\linewidth]{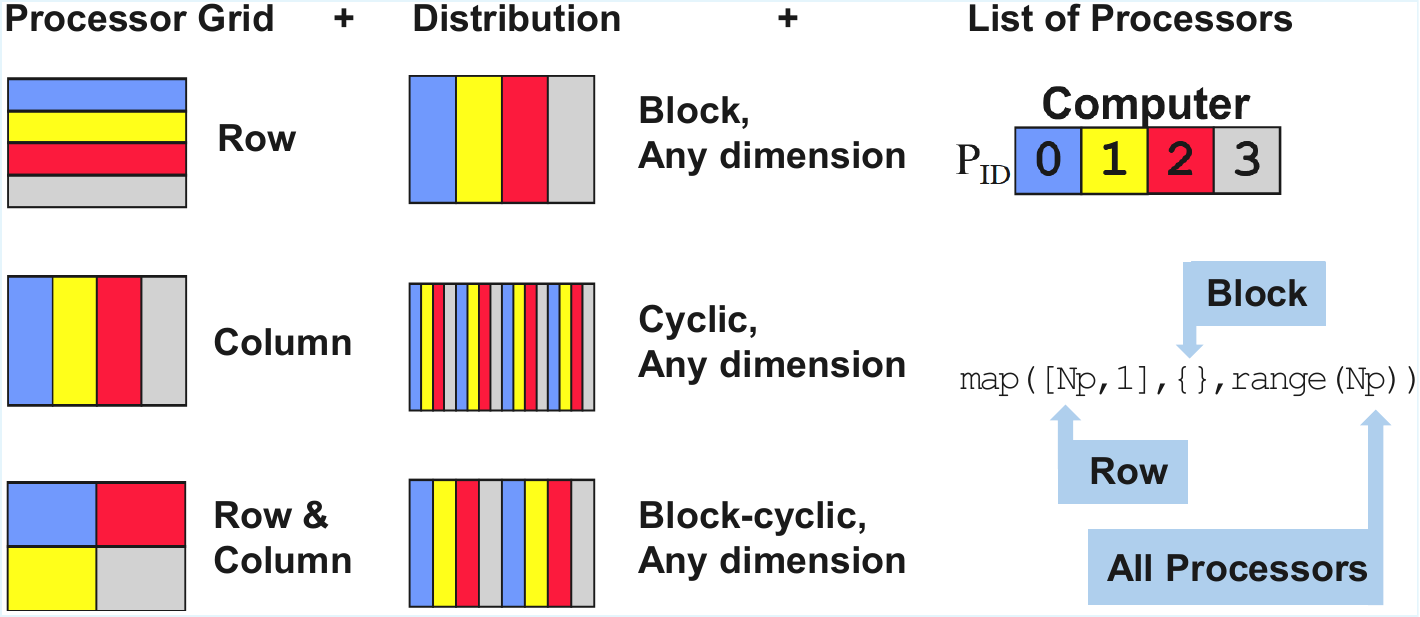}
    \caption{Maps are introduced to handle how work is distributed among processes. A map has three main parameters: the processor grid, the distribution, and the list of processors. The processor grid makes a grid that is applied to the array, and allows the array to be divided in rows, columns, or rows and columns. Distribution refers to how pieces of the array are distributed to the processors, and can be specified as block, cyclic, block-cyclic, and block-overlap (not pictured). The list of processors describes which processors will receive a part of the distributed array created from this mapping. The map used for benchmarking in this paper divides the rows in blocks, and distributes work to all processors.}
    \label{fig:MapDiagram}
\end{figure}

More specifically, while the {\tt TrafficMatrix} class might be useful in broader contexts, it added unnecessary overhead for the specific Graph Challenge use case. Each class instance kept excessive information on each time window once subranges were involved, for example, creating and keeping track of all the resulting subrange-masked matrices past when they were most useful. This has the potential for memory issues when dealing with larger amounts of data. Additionally, separate functions handled analysis computations for subrange matrices despite the mathematical equivalence of the underlying matrix operations. Multiple optional features obscured the essential algorithm structure, making it difficult for new researchers to understand the core approach.  

The streamlined implementation consists of two concise files totaling ~325 lines of code:
\begin{enumerate}
    \item {\bf Tar \& Matrix Processing ($\sim85$ lines)}: Functions for tar file management and GraphBLAS matrix processing. 
    \item {\bf Challenge Pipeline ($\sim240$ lines)}: Higher-level functions for summation, analysis, and processing, including the main {\tt process\_filelist} function that completes the full step for a singular time window.
\end{enumerate}

Additional abstractions were removed in favor of direct GraphBLAS matrix operations, especially for analysis. A single analysis function now handles all matrix variants and computes all nine properties described in Table 1 found in the Anonymized Network Sensing Graph Challenge Paper \cite{jananthan2024anonymized} altogether, reusing relevant values. This implementation follows a clear sequential pattern that reflects the basic scenario of the Anonymized Network Sensing Graph Challenge, providing a more accessible foundation for community participation and allowing researchers to easily display their improvements.

\begin{figure}
    \centering
\begin{CodeListing}[{\bf Parallel Matlab program}]
\label{code:ParallelMatlab}
\begin{codebox}
\\
\li
\li {\tt\footnotesize \% filelist is a list of tar file names to process.}
\li {\tt\footnotesize N = length(filelist)}
\li {\tt\footnotesize Filemap = map([Np,1],\{\},0:Np-1);}  \>\>\>\>\>\>\>\>\>\>\>{\tt\footnotesize \%  Map.}
\li {\tt\footnotesize z = zeros(N, 1, map=Filemap);}
\li {\tt\footnotesize my\_i\_global = global\_ind(z,1);}		
\li {\tt\footnotesize for i\_local = 1:length(my\_i\_global)}
\li \> {\tt\footnotesize     i\_global = my\_i\_global(i\_local);}
\li \> {\tt\footnotesize     file = filelist\{i\_global\};}
\li \> {\tt\footnotesize \% Continue to read, sum, analyze this file}
\li {\tt\footnotesize end}
\end{codebox}
\end{CodeListing}
\end{figure}

\section{Distributed Arrays for Parallelization}
 
The streamlined serial reference implementation was benchmarked by using pPython ~\cite{byun2023ppython} and pMatlab ~\cite{kepner2009parallel} to evaluate performance on parallel architectures. Both libraries use map-based distributed array programming concepts to partition data across multiple processors while keeping the serial algorithm clear and intact. 

Maps provide an abstraction that separates the algorithm design from parallelization. A map defines three key elements that describe how an array will be distributed between processes, shown in Figure~\ref{fig:MapDiagram}. The processor grid describes how to section the array, which can be in any dimension, like rows or columns, or can be multiple dimensions. Distribution specifies how the array is distributed between processes, and can be in block, cyclic, block-cyclic, or block-overlap in manner. The list of processors which describes which processors will receive data distributed by the created map.

Code excerpts for benchmarking Graph Challenge implementations that read, sum, and analyze matrices using distributed arrays in Matlab/Octave and Python are shown in Code Listings~\ref{code:ParallelMatlab} and \ref{code:ParallelPython}.  The implementations use pMatlab  and pPython to distribute work among processes, the amount of which is denoted by $N_P$.  

The method works as follows. First, a map is created to distribute work among processors. The map divides equally the rows of the array, with {\tt \{\}} indicating that it is distributed in the default block distribution, and all processors, denoted by a $P_{ID}$, from $P_{ID}$ = 0 to $P_{ID}$ =  $N_P-1$ are included. An empty array structure is also initialized, with the first dimension being the amount of file time windows that must be processed in total. This will represent the work that needs to be done, and by tracking the indices that are split between processes, each unique $P_{ID}$ will know what section of work it is responsible for. See Figure~\ref{fig:MapDiagram} for a visualization of how the map was created and options available when creating a map.

All processes execute the code found in Code Listings~\ref{code:ParallelMatlab} and~\ref{code:ParallelPython}, but each operates on its locally assigned files. Because each file can be processed independently, no inter-process communication is required, eliminating any communication or synchronization overhead.

\begin{figure}    
\begin{CodeListing}[{\bf Parallel Python program}]
\label{code:ParallelPython}
\begin{codebox}
\\
\li {\tt\footnotesize import pPython as GPC}
\li {\tt\footnotesize \# filelist is a list of tar file names to process.}
\li {\tt\footnotesize N = len(filelist)}
\li {\tt\footnotesize Filemap = Dmap([Np,1],\{\},range(Np))}  \>\>\>\>\>\>\>\>\>\>\> {\tt\footnotesize \#  Map.}
\li {\tt\footnotesize z = zeros(N, 1, map=Filemap)}
\li {\tt\footnotesize my\_i\_global = global\_ind(z,0)[0]}
\li {\tt\footnotesize for i\_local in range(len(my\_i\_global)):}
\li \> {\tt\footnotesize     i\_global = my\_i\_global[i\_local]}
\li \> {\tt\footnotesize     file = filelist[i\_global]}
\li \> {\tt\footnotesize \# Continue to read, sum, analyze this file}
\li
\end{codebox}
\end{CodeListing}

\end{figure}
The descriptions of Code Listings \ref{code:ParallelMatlab} and \ref{code:ParallelPython} are as follows:

\noindent {\bf Line 1} of the Python code imports the required Python package. {\bf Line 4} creates the parallel map {\tt Filemap} specifying that the $N{\times}1$ element column vector is to have its rows divided equally using the default block distribution (\{\}) among all {\tt Np} over processes with $P_{ID}$ = {\tt 0}, ..., {\tt Np-1}.

This benchmarking script showcases the strengths of the distributed array programming approach. 
\begin{itemize}
\item {\bf Low Code Impact}. Only a few additional lines of code are necessary to convert the serial program into a parallel program. 
\item {\bf Small Parallel Library Footprint}. Only two parallel library functions were required: {\tt map} and {\tt global\_ind}.
\item {\bf Scalable}. The code can be run on any problem size or number of processors such that $N > N_P$.
\item {\bf Map Independence}. The program will work for any distribution (i.e., block, cyclic, block-cyclic).
\item {\bf Performance Guarantee}. Because any following operations are working on a separate file that is local to the processor, there is a guarantee that there is no hidden performance penalty when running these lines of code. 
\end{itemize}

Memory constraints were addressed by partitioning workloads into smaller time windows when benchmarking, ensuring that each process handled a manageable subset of data. Code listing for the parallel Matlab/Octave and Python implementations are provided in Code Listings~\ref{code:ParallelMatlab} and~\ref{code:ParallelPython}.  

\section{Performance Measurements}
Benchmark measurements were taken using dual 2.4 GHz Intel Xeon Platinum 8260 processor compute nodes on the MIT SuperCloud TX-E1 supercomputer \cite{reuther2018interactive}. The refactored code was tested using Python, Matlab, and Octave, with  pMatlab distributed arrays used for Matlab/Octave, and pPython similarly for Python. Starting with a singular thread on one process and one node, threads were doubled until 16. Then multi-node benchmarking was completed, working with 16 threads per process, three processes per node, starting with a singular node and doubling nodes until 32. 

\begin{figure}
\center{\includegraphics[width=1.0\columnwidth]{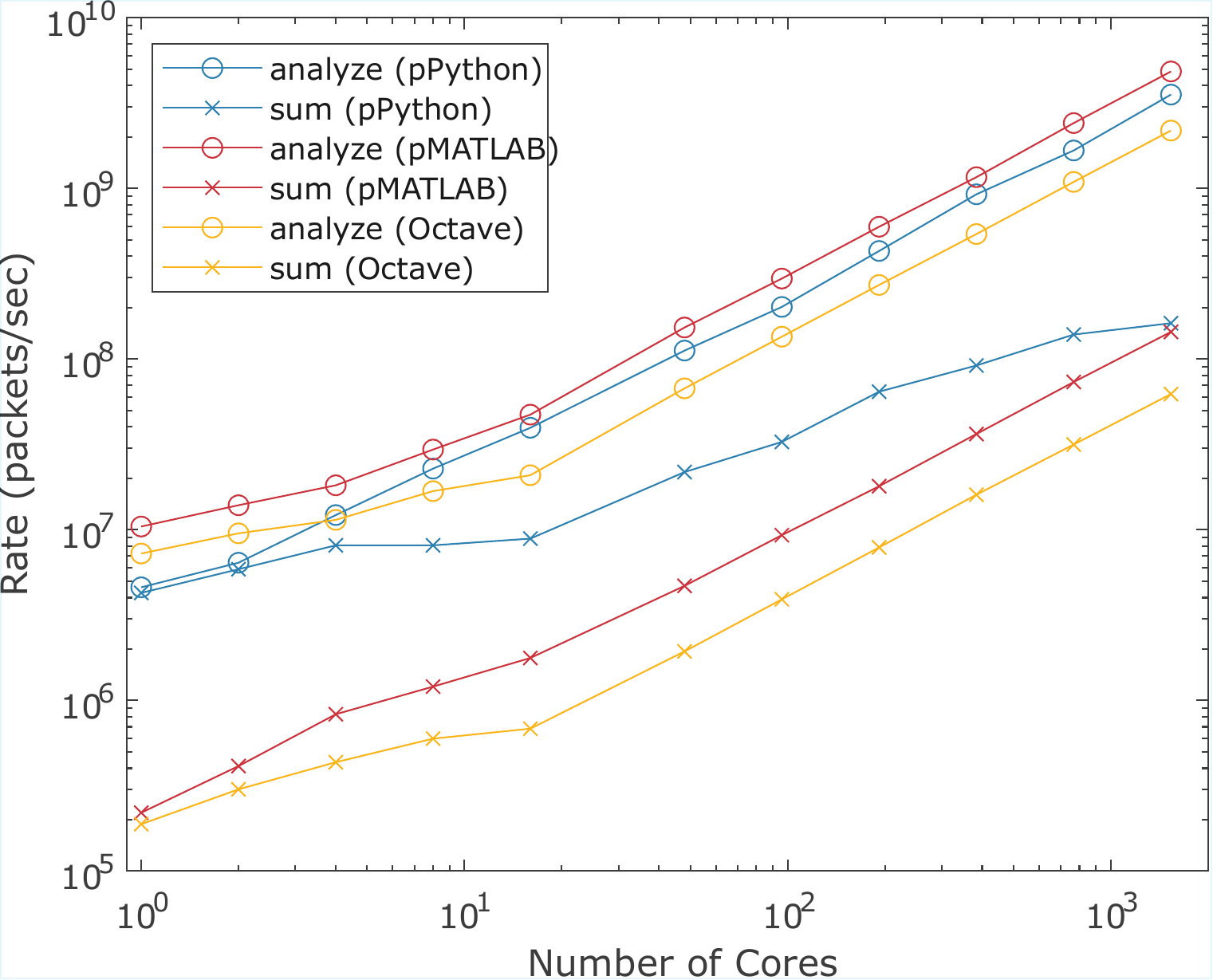}}
	\caption{{\bf Performance Comparison.} Average performance of traffic matrix sum and analyze code across three languages - Python, MATLAB, and Octave. pMATLAB distributed arrays were used to divide files among 3 instances, each with 16 OpenMP threads. The sum and analyze code have a larger memory footprint and 3 distinct instances each with 16 OpenMP threads each processing a separate file can be run on a 48 core node with 192 GB of RAM.  The multi-node performance scales linearly over distinct files.}
	\label{fig:ReferenceLangPerformance}
\end{figure}

Performance results comparing the Python, Matlab, and Octave implementations are shown in Figure~\ref{fig:ReferenceLangPerformance}. Sum and analysis of the traffic matrices requires a larger memory footprint, which can be accelerated with threads.  In both cases, multiple files can be processed simultaneously, and the performance scales linearly with nodes.

Key observations include:
\begin{itemize}
    \item Summation consistently required more time than analysis across all languages. 
    \item  Python summation outperformed Matlab and Octave, likely due to optimized in-place addition. Analysis times were similar across languages as it directly tests the GraphBLAS capabilities. 
    \item Performance scaled nearly linearly with increasing cores and nodes, with slight drop-offs for Python in summation at larger scales. 
\end{itemize}

These findings highlight the trade-offs between languages and GraphBLAS capabilities. Distributed array programming is shown to be well-suited for parallelizing this type of problem, making it low effort to achieve speed-ups in file processing across more nodes.

\section{Conclusions}
This work presents a refactoring of the last step of the Anonymized Netowrk Sensing Graph Challenge reference implementation that achieves a 67\% reduction in code size while maintaining full functionality. The simplification improved clarity and adaptability by removing abstractions and redundant computations, making the implementation a more effective community reference. This work makes a more accessible and efficient reference implementation for the Graph Challenge community, lowering barriers for new researchers to participate in the challenge. 

Distributed array techniques and the additions of maps effectively distributed work among processes while keeping the code impact low. The benchmarking study shows the ease of parallel additions to scale performance linearly when utilizing additional nodes and can provide information into language and tool trade-offs.

\section*{Acknowledgments}

The authors wish to acknowledge the following individuals
for their contributions and support: Daniel Andersen, LaToya Anderson, Bill Arcand, Sean Atkins, Bill Bergeron, Chris Berardi, Bob Bond, Alex Bonn, Bill Cashman, K Claffy, Tim Davis, Chris Demchak, Alan Edelman, Peter Fisher, Jeff Gottschalk, Thomas Hardjono, Chris Hill, Michael Houle, Michael Jones, Charles Leiserson, Piotr Luszczek, Kirsten Malvey, Peter Michaleas, Lauren Milechin, Chasen Milner, Sanjeev Mohindra, Guillermo Morales, Julie Mullen, Heidi Perry, Sandeep Pisharody, Christian Prothmann, Andrew Prout, Steve Rejto, Albert Reuther, Antonio Rosa, Scott Ruppel, Daniela Rus, Mark Sherman, Scott Weed, Charles Yee, Marc Zissman.

\bibliographystyle{ieeetr}
\bibliography{ImproveGraphChal}

\end{document}